# The Calculus of Expected Loss:

# Backtesting Parameter-Based Expected Loss in a Basel II Framework

Wolfgang Reitgruber[1][2][3]

**Abstract**

The dependency structure of credit risk parameters is a key driver for capital consumption and receives regulatory and scientific attention. The impact of parameter imperfections on the quality of expected loss (EL) in the sense of a fair, unbiased estimate of risk expenses however is barely covered. So far there are no established backtesting procedures for EL, quantifying its impact with regards to pricing or risk adjusted profitability measures.

In this paper, a practically oriented, top-down approach to assess the quality of EL by backtesting with a properly defined risk measure is introduced. In a first step, the concept of risk expenses ("Cost of Risk") has to be extended beyond the classical provisioning view, towards a more adequate capital consumption approach ("Impact of Risk", IoR). On this basis, the difference between parameter-based EL and actually reported "Impact of Risk" is decomposed into its key components.

The proposed method will deepen the understanding of practical properties of EL, reconciles the EL with a clearly defined and observable risk measure and provides a link between upcoming IFRS 9 accounting standards for loan loss provisioning with IRBA regulatory capital requirements. The method is robust irrespective whether parameters are simple, expert based values or highly predictive and perfectly calibrated IRBA compliant methods, as long as parameters and default identification procedures are stable.

---

[1] Department for Strategic Risk Management, UniCredit Bank Austria AG,
Julius Tandlerplatz 3, A-1090 Vienna, Austria; email: wolfgang.reitgruber@unicreditgroup.at
The presented opinions and methods do not necessarily correspond to those of UniCredit Bank Austria AG.

## 1. Introduction and motivation

For financial institutions reporting under IRBA approach in Basel II framework, internal steering processes (eg pricing for risk expenses or cost center accounting) widely apply EL = PD*EAD*LGD as key measure for cost of risk. It is the standard measure to fulfill regulatory use-test conditions, deviations have to be justified.

The product of the credit risk parameters PD, EAD and LGD however is statistically challenging. In scientific literature, few articles address this question: Some highlight the evidence of correlations between PD and LGD in certain markets, eg Altman *et al* (2005), Altman (2006) or Farinelli *et al* (2012). Other publications try to propose certain models to adjust credit risk models for these dependencies, such as Gaspar and Slinco (2008) or Wan and Dev (2007). Some are related to the impact of PD-LGD correlation on unexpected loss, eg Meng *et al* (2010) or Miu and Ozdemir (2005, 2006). In spite of the significant methodological risk caused by these dependencies or any other methodological imperfections of PD, EAD or LGD models, a practical quantification of the impact to EL estimation through backtesting with reported, incurred risk is not yet established.

To mitigate this methodological risk, the Basel II framework on IRBA models imposes certain conditions related to the dependency structure of PD, EAD and LGD: “Credit institutions shall use LGD estimates that are appropriate for an economic downturn”. “In case of dependency between risk of obligor and risk of collateral, conservative measures are to be applied”. “… the conversion factor estimate shall incorporate a larger margin of conservativity where a stronger positive correlation can reasonably be expected between the default frequency and the magnitude of the conversion factor”. Besides being methodologically less strict than statistical independency, it may cause overly conservative EL estimates. Current empirical research (eg Samuels (2013), Cannata *et al* (2012) and PECDC (2013)) indicates that, although there are strict validation requirements under IRBA, the resulting RWA figures even for comparable portfolios may differ significantly. It seems that there is still a large degree of inconsistency possible in building models for credit risk.

Another issue for practical application is related to the outcome period: For PD and EAD parameters, a 12-month outcome period is prescribed. The LGD parameter is inherently a lifetime value. Consequently, backtests for PD and EAD parameters are typically based on a finite 12-month observation window, while mainstream LGD backtests are generally based on a long-term vintage approach. The EL is somewhere in between: partly a 12-month measure (and treated in internal steering for performing loans this way) and partly a lifetime measure (as soon as LGD or loan loss provisions are concerned). This inconsistency in the definition of the 12-month EL already caused controversial discussions on its methodological validity during review phase of IFRS 9 in some countries. There is currently no established industry standard EL backtest based on a finite 12-month observation period.

What makes matters worse, many businesses may observe significant differences between EL and actually reported risk expenses in some segments. In spite of strong pressure to apply the EL for internal steering through the use test requirement, it is for this reason often regarded an analytical concept, of limited practical relevance. The situation can be compared to the allocation procedures of operational expenses: total expenses are easily visible in the P&L, which are to be allocated to business segments for detailed profitability assessments. The total is easily reconciled, and the allocation discussion can focus on distribution parameters. The EL has no such counterpart on total bank level (reconciled with banks accounts), although it is effectively allocated to each customer account through pricing models or in risk adjusted profitability measures. There is clearly the need for a better alignment of the total EL to a measure of risk expenses which is relevant and visible for top-management.

Comment: with the upcoming IFRS 9 accounting standards for loan loss provisioning, the methodological gap between accounting and regulation will be significantly reduced. Provisions for performing loans shall be based on EL measures, although some differences in detail are to be expected (eg IFRS 9 is going to be more biased towards procyclical PIT/fair outlook estimates instead of regulatory anticyclical TTC type estimates, the consideration of lifetime EL instead of 12-month EL in higher risk segments, to name a few). Depending on the specific details of the final accounting standards, the proposed concept of “Impact of Risk” (defined below) can be generalized to derive an EL backtest directly compatible with P&L impact.

In this paper, a measure of appropriateness of the EL based on actual risk impact on capital will be derived. The quality (or bias) of the EL will be quantified irrespective of the quality of underlying parameters and without requiring independency between parameters. In spite of the hybrid nature of the EL being partly focused on 12 months outcome, partly on lifetime loss prediction, the proposed backtest requires only 12 months of data and does not need long term vintage analyses to be applied. The necessary data (basically the EL at begin and end of an accounting period split by performing and non-performing segments, and write-off amounts during that period) are generally available for financial institutions reporting under IRBA and are easily reconciled on business level. This leads us back to the “operational expense” example: a reconciled risk measure gets allocated to customer accounts and business segments through the EL and the results of the respective backtests.

The proposed measure quantifies the bias of the EL estimate and attributes it to surplus/shortage of defaults or surplus/shortage of recoveries. It supports a detailed discussion of outliers, cyclical effects and general, methodologically driven deviations (conservativities, discounting effects or downturn assumptions in LGD). The calculus is fairly straightforward and should lead analytically minded managers to a deeper understanding of EL and risk impact on capital. It is a perfect starting point to identify critical risk segments or refine credit risk parameters. Understanding the EL through this framework will improve the credit risk budgeting process and will help to explain risk performance during monthly, quarterly or annual performance reviews. In this sense the target audience goes definitely beyond the analytical teams – whoever needs or wants a deeper understanding of risk impact on capital and who plans to use this knowhow to improve the quality of strategic decisions will find this article helpful.

Key references can be found in the general regulatory framework around Basel II. There is extensive literature available for credit risk modeling generally, eg McNeil *et al* (2005) or Benzschawel (2012). Validation and backtesting textbooks are mostly focusing on single factors, eg Miu and Ozdemir (2008) or Engelmann and Rauhmeier (2006, 2011). Meyer and Quell (2011) summarize the current industry standard: statistical backtesting concepts for QRM for credit risk are almost impossible as P&L-derived actual risk performance data are inherently annual figures. There are simply insufficient data available for meaningful backtest analyses. Many articles work on “point-in-time” versus “through-the-cycle” concepts, mostly related to the PD-LGD relationship and the requirement of downturn LGD in Basel II, eg Altman *et al* (2005, 2006), Gaspar and Slinco (2008), Meng *et al* (2010), Miu and Ozdemir (2005, 2006) or Wan and Dev (2007). To follow up on the relationship of EL with pricing you may refer to the reference volume by Duffie and Singleton (2003) or Berg (2010). So far nothing close to the presented work on EL backtesting seems to be published already.

## 2. Definition of “Impact of Risk”

Which types of risk measures can be related to expected loss? Traditionally, loan loss provisions and write-offs are the main P&L drivers for credit risk. This is fully in line with the “incurred loss” approach in classical accounting frameworks until IAS 39: after a certain loss indication is observed, a provision has to be created. The following expression for risk expenses related to loan loss provision (LLP) is easily derived:

(1) $$\mathrm{LLP}^{\mathrm{EOP}} = \mathrm{LLP}^{\mathrm{BOP}} - \text{LLP usage} + \text{new generated LLP} - \text{LLP releases}$$

$$\begin{aligned}\text{LLP Expenses} &= \text{new generated LLP} - \text{LLP releases} + \text{direct write off} = \\ &= \mathrm{LLP}^{\mathrm{EOP}} - \mathrm{LLP}^{\mathrm{BOP}} + \text{LLP usage} + \text{direct write off}\end{aligned}$$

(2) $$\text{LLP Expenses} = \mathrm{LLP}^{\mathrm{EOP}} - \mathrm{LLP}^{\mathrm{BOP}} + \text{total write off}$$

Throughout this paper EOP is used for "End of Period", typically the end of an accounting period. BOP stands for "Begin of Period", typically corresponding to the end of the previous accounting period, one year earlier. LLP usage is write-off against existing provisions, direct write-offs occur in case there is no existing provision booked yet. Provisions will be released if the customer recovers, new provisions are generated if the credit quality is deteriorating.

With IAS 39, the IBNR ("Incurred but Not Reported") concept introduces provisions on the performing loan portfolio as well, through the argument that these losses "already occurred but are not yet detected". IBNR provisions are typically based on EL on performing loan portfolio by multiplication with the loss confirmation period (ie *6/12 in case of a 6 month loss confirmation period). As there is no "usage" of this type of provision through write-offs, any changes impact P&L:

(3) $\text{IBNR Expenses} = \text{IBNR}^{\text{EOP}} - \text{IBNR}^{\text{BOP}}$

With the introduction of Basel II together with an increasing focus of regulators on capital requirements, risk impact on capital became at least as important for financial institutions as risk impact on P&L. Whether available capital is used to fund provisions or a shortfall has to be considered when minimum (regulatory) capital requirements are assessed is equivalent with respect to potential shareholder return. Current regulatory trends with increasing minimum capital ratios and stronger quality requirements on available capital (eg in Basel III) further stress this development. Increased visibility and relevance of EBA stresstest results have an impact as well. The enforcement of the JRAD process (CEBS (2010)) with respect to actual regulatory capital requirement also underlines the focus on capital consumption.

Besides LLP and IBNR, the remaining risk impact on capital under IRBA framework is given by the shortfall. Financial institutions are requested to deduct it from available Tier 1 and Tier 2 capital: "Banks need to compare the IRB measurement of expected losses (EAD*PD*LGD) with the total amount of provisions … Shortfall amounts must be deducted from capital … 50% from Tier 1, 50% from Tier 2. Excess amounts are eligible as Tier 2 capital … limited by supervisory discretion."

The shortfall (SF) is defined by

(4) $\text{SF}^{\text{EOP}} = \text{EL}^{\text{EOP}} - \text{LLP}^{\text{EOP}} - \text{IBNR}^{\text{EOP}}$

and its impact on capital during one accounting period is

(5) $\text{SF Impact to Capital} = \text{SF}^{\text{EOP}} - \text{SF}^{\text{BOP}}$

Comment: formally, this shortfall may also become negative, ie representing an excess. It depends on local regulatory discretions, whether this situation can be treated symmetrically with regard to capital requirements or not. Because of the need for conservativity in most EL estimates (through conservative, partially downturn-based parameters) and loss detection periods in the estimation of the IBNR of mostly below 12 months, a (positive) shortfall is generally to be expected. Although there are some local regulations which may limit the deductibility of IBNR from the applicable shortfall, the core concept of this paper remains completely valid.

Comment 2: This Basel 2 shortfall concept effectively provides the link between the regulatory-based EL and accounting measures such as loan loss provisions for non-performing loans and IBNR for performing loans under IAS 39. With the upcoming introduction of IFRS 9, alignment between regulation and provisioning with respect to the use of parameters for credit risk measurement will become real, allowing for a more consistent, methodological treatment of the shortfall. As immediate consequence, changes to EL will not only impact capital requirements (current status), but have a corresponding impact on P&L as well. Based on the overall positive appreciation of the Basel committee of the latest exposure draft on IFRS 9 in Basel (2013), further convergence between regulation and accounting can be expected.

Direct and indirect workout cost can be considered in the proposed concept through LGD quite naturally. They will not be considered explicitly in this paper to keep the framework simple.

Now the term “Impact of Risk” can be introduced as the combined impact of loan loss provisions (LLP and IBNR) and shortfall:

(6) $\mathrm{IoR} := \Delta\mathrm{LLP} + \text{total write off (wo)} + \Delta\mathrm{IBNR} + \Delta\mathrm{SF}$

which can be transformed to

Lemma (7) $\mathrm{IoR} = \mathrm{EL}^{\mathrm{EOP}} - \mathrm{EL}^{\mathrm{BOP}} + \mathrm{wo}$

Proof of lemma (7):

$$\begin{aligned}\mathrm{IoR} = {} & \mathrm{LLP}^{\mathrm{EOP}} - \mathrm{LLP}^{\mathrm{BOP}} + \mathrm{wo} + \mathrm{IBNR}^{\mathrm{EOP}} - \mathrm{IBNR}^{\mathrm{BOP}} + \\ & +\left(\mathrm{EL}^{\mathrm{EOP}} - \mathrm{LLP}^{\mathrm{EOP}} - \mathrm{IBNR}^{\mathrm{EOP}}\right) - \left(\mathrm{EL}^{\mathrm{BOP}} - \mathrm{LLP}^{\mathrm{BOP}} - \mathrm{IBNR}^{\mathrm{BOP}}\right) = \\ = {} & \mathrm{EL}^{\mathrm{EOP}} - \mathrm{EL}^{\mathrm{BOP}} + \mathrm{wo} \qquad \text{q.e.d.}\end{aligned}$$

Basically, the EL (covering the whole portfolio – performing loans as well as non-performing loans) takes the role of LLP in this framework. Any change in EL occurring during the observation period immediately causes IoR to increase or decrease, unless caused by a write-off. Write-offs (direct or through LLP usage) reduce EL, with low residual impact as long as the LGD is approaching 100% of exposure before write-off. For further reading please follow up with question 1 in the appendix.

Objectivity of timing is the key factor in this process: Write-off procedures may differ significantly from country to country and from segment to segment (eg retail versus corporate, mortgage versus credit card). It is mostly an accounting-driven process, sometimes preferring extremely late write-off to reasonably early write-off to ensure that any remaining recoveries get collected properly.

Loan loss provisions are also - to a large extent - driven by rule-based manual processes. Timing is focused on accounting periods, leading to increased provisioning activity during the last fiscal quarter. As such, provision-based cost-of-risk measures contain a significant amount of subjective timing and amount decisions and might even amplify cyclical credit risk patterns (worst case).

Consistency and calibration of provisioning and write-off processes are difficult to assess methodologically. Consequently, annual LLP movements remain questionable as unbiased (and uncorrelated through time) measures of annual credit risk.

Impact of Risk (IoR) provides an objective measure to assess credit risk performance for a certain period, with immediate business relevance for IRBA portfolios due to its direct interpretation through "usage of available capital". Timing is prescribed mostly by default identification, amounts by EAD and LGD parameters, exposure movements and collateral re-evaluations. With stable and objective default identification, rating updating and collateral re-evaluation procedures, it can be expected that IoR provides a statistically viable monthly measure for credit risk.

This section can be closed by summarizing the following observations:

a) Whenever write-offs occur, EL should already be close to 100% of exposure. This is a consequence of adequate LGD Best Estimate models approaching 100% short before the write-off is actually realized. As a consequence, IoR is robust with respect to timing of write-off.

b) Key driver for IoR is the default identification process, which constitutes the most significant discontinuity in the EL due to the change from an effectively 12 month outcome period (before default, EL based on 1-year PD) to lifetime loss (after default, EL being represented by the lifetime LGD) and the corresponding increase of PD from mostly single digit percentages to 100%. Since default identification is highly regulated in the Basel II framework, a much higher degree of objectivity compared to loan loss provisioning- or write-off-processes can be assumed. Nevertheless additional analyses should be performed to compare the stability and other statistical qualities of alternative actual credit risk measures.

c) No assumptions are necessary on the quality of EL in terms of calibration or predictive power of its components, besides stability of the underlying PD, EAD and LGD models: an overly conservative EL leads to earlier/frontloaded credit risk recognition, whereas an aggressive EL estimate would cause a delay, compared to the situation with a properly calibrated EL. Besides this time shift (which would cause an overall impact when discounting for net present values is considered), the overall credit risk amount over life time does not change irrespective of the EL calibration. Question 1 in the appendix explains this relationship by means of a simple example. Application of lifetime EL in possible future accounting standards (eg IFRS 9) makes no difference: risk recognition will happen even earlier (ie increasing conservativity), but the overall credit risk amount over lifetime would not change either.

d) Although the proposed concept is derived under IRBA considerations with standard local regulatory discretions (with respect to shortfall and IBNR deductibility), it is methodologically applicable to every portfolio. Differences between IoR and actually incurred impact on capital caused by local discretions (regulations or accounting standards) can be easily quantified.

**3. The decomposition of Impact of Risk (IoR)**

Theorem (8): $\mathrm{IoR} = \mathrm{EL}_{\mathrm{PL}}^{\mathrm{EOP}} + \mathrm{PL\ Backtest} + \mathrm{NPL\ Backtest}$

With the terms

(9) PL Backtest $:= \mathrm{EL}_{\mathrm{new\ NPL}}^{\mathrm{EOP}} + \mathrm{wo}_{\mathrm{new\ NPL}} - \mathrm{EL}_{\mathrm{PL}}^{\mathrm{BOP}}$

(10) NPL Backtest $:= \mathrm{EL}_{\mathrm{old\ NPL}}^{\mathrm{EOP}} + \mathrm{wo}_{\mathrm{old\ NPL}} - \mathrm{EL}_{\mathrm{NPL}}^{\mathrm{BOP}}$

Proof of theorem (8): EL is split by performing loans (PL) and non-performing or impaired loans (NPL). The NPL segment at EOP is further split by “new NPL” and “old NPL” to differentiate, whether a default occurred between BOP and EOP or already before:

$$\mathrm{IoR} = \mathrm{EL}_{\mathrm{PL+NPL}}^{\mathrm{EOP}} - \mathrm{EL}_{\mathrm{PL+NPL}}^{\mathrm{BOP}} + \mathrm{wo} =$$

$$= \mathrm{EL}_{\mathrm{PL}}^{\mathrm{EOP}} + \mathrm{EL}_{\mathrm{new\ NPL}}^{\mathrm{EOP}} + \mathrm{EL}_{\mathrm{old\ NPL}}^{\mathrm{EOP}} - \mathrm{EL}_{\mathrm{PL}}^{\mathrm{BOP}} - \mathrm{EL}_{\mathrm{NPL}}^{\mathrm{BOP}} + \mathrm{wo}_{\mathrm{new\ NPL}} + \mathrm{wo}_{\mathrm{old\ NPL}} =$$

$$= \mathrm{EL}_{\mathrm{PL}}^{\mathrm{EOP}} + \left(\mathrm{EL}_{\mathrm{new\ NPL}}^{\mathrm{EOP}} + \mathrm{wo}_{\mathrm{new\ NPL}} - \mathrm{EL}_{\mathrm{PL}}^{\mathrm{BOP}}\right) + \left(\mathrm{EL}_{\mathrm{old\ NPL}}^{\mathrm{EOP}} + \mathrm{wo}_{\mathrm{old\ NPL}} - \mathrm{EL}_{\mathrm{NPL}}^{\mathrm{BOP}}\right) =$$

$$= \mathrm{EL}_{\mathrm{PL}}^{\mathrm{EOP}} + \mathrm{PL\ Backtest} + \mathrm{NPL\ Backtest} \qquad \text{q.e.d.}$$

IoR is now effectively decomposed into three components:

a) $\mathrm{EL}_{\mathrm{PL}}^{\mathrm{EOP}}$: The EL risk premium incurred by ongoing performing transactions and new transactions.

b) PL Backtest = $\mathrm{EL}_{\mathrm{new\ NPL}}^{\mathrm{EOP}} + \mathrm{wo}_{\mathrm{new\ NPL}} - \mathrm{EL}_{\mathrm{PL}}^{\mathrm{BOP}}$: The positive or negative deviation of new defaults from EL prediction at begin of period.

c) NPL Backtest = $\mathrm{EL}^{\mathrm{EOP}}_{\mathrm{old\ NPL}} + \mathrm{wo}_{\mathrm{old\ NPL}} - \mathrm{EL}^{\mathrm{BOP}}_{\mathrm{NPL}}$: The positive or negative deviation of recoveries compared to expectations by the LGD model dynamic on the NPL portfolio. Please note that only LGD dynamic remains relevant for the NPL Backtest, since PD as well as EAD are already realized.

Observation: $\mathrm{EL}^{\mathrm{EOP}}_{\mathrm{PL}}$ is not the obvious term one would expect in this situation – IoR during one period to be driven by the end-of-period rating/portfolio distribution and two error terms (backtests) may seem to be a surprising result. Common sense would expect $\mathrm{EL}^{\mathrm{BOP}}_{\mathrm{PL}}$ instead. This replacement is caused by the Basel II IRBA shortfall regulation, requesting $\mathrm{EL}_{\mathrm{PL}}$ to be considered in the capital requirement, even though it is just a prediction and does not correspond to realized credit risk during the accounting period yet. In other words: Basel II IRBA regulation formally requires frontloading of credit risk recognition by one year, caused by the 12-month forecasting horizon in the EL estimation.

Alternative 1: Would the shortfall definition be based on $\mathrm{EL}_{\mathrm{NPL}}$ instead of $\mathrm{EL}_{\mathrm{PL+NPL}}$, $\mathrm{IoR}_{\mathrm{NPL}} = \mathrm{EL}^{\mathrm{BOP}}_{\mathrm{PL}} + \mathrm{PL\ Backtest} + \mathrm{NPL\ Backtest}$ can be easily derived for the correspondingly re-defined $\mathrm{IoR}_{\mathrm{NPL}}$ measure. $\mathrm{IoR}_{\mathrm{NPL}}$ is neutral with respect to timing credit risk recognition (compared to income recognition through risk margin), fully aligned with default identification.

Lemma (11): $\mathrm{IoR}_{\mathrm{NPL}} = \mathrm{EL}^{\mathrm{BOP}}_{\mathrm{PL}} + \mathrm{PL\ Backtest} + \mathrm{NPL\ Backtest}$

Proof of lemma (11):

$$\mathrm{IoR}_{\mathrm{NPL}} = \mathrm{EL}^{\mathrm{EOP}}_{\mathrm{NPL}} - \mathrm{EL}^{\mathrm{BOP}}_{\mathrm{NPL}} + \mathrm{wo} = \left(\mathrm{EL}^{\mathrm{EOP}}_{\mathrm{PL+NPL}} - \mathrm{EL}^{\mathrm{BOP}}_{\mathrm{PL+NPL}} + \mathrm{wo}\right) - \left(\mathrm{EL}^{\mathrm{EOP}}_{\mathrm{PL}} - \mathrm{EL}^{\mathrm{BOP}}_{\mathrm{PL}}\right) =$$

$$= \left(\mathrm{EL}^{\mathrm{EOP}}_{\mathrm{PL}} + \mathrm{PL\ Backtest} + \mathrm{NPL\ Backtest}\right) - \left(\mathrm{EL}^{\mathrm{EOP}}_{\mathrm{PL}} - \mathrm{EL}^{\mathrm{BOP}}_{\mathrm{PL}}\right) =$$

$$= \mathrm{EL}^{\mathrm{BOP}}_{\mathrm{PL}} + \mathrm{PL\ Backtest} + \mathrm{NPL\ Backtest} \qquad \text{q.e.d.}$$

Alternative 2: Replacing $\mathrm{EL}_{\mathrm{PL}}$ completely or partly by lifetime EL, the correspondingly re-defined term $\mathrm{IoR}_{\mathrm{life_PL}} = \mathrm{EL}^{\mathrm{EOP}}_{\mathrm{life_PL+NPL}} - \mathrm{EL}^{\mathrm{BOP}}_{\mathrm{life_PL+NPL}} + \mathrm{wo}$ represents the methodological counterpart in line with current drafts of accounting standards for expected credit losses as developed by the FASB or IASB. In this case, even faster acceleration of credit risk recognition can be observed, leading to more conservativity (ie creation of hidden reserves) compared to income recognition.

Lemma (12): $\mathrm{IoR}_{\mathrm{life_PL}} = \left(\mathrm{EL}^{\mathrm{EOP}}_{\mathrm{PL}} + \mathrm{PL\ Backtest} + \mathrm{NPL\ Backtest}\right) + \left(\mathrm{EL}^{\mathrm{EOP}}_{\Delta\mathrm{PL}} - \mathrm{EL}^{\mathrm{BOP}}_{\Delta\mathrm{PL}}\right) =$

$$= \left(\mathrm{EL}^{\mathrm{BOP}}_{\mathrm{PL}} + \mathrm{PL\ Backtest} + \mathrm{NPL\ Backtest}\right) + \left(\mathrm{EL}^{\mathrm{EOP}}_{\mathrm{life_PL}} - \mathrm{EL}^{\mathrm{BOP}}_{\mathrm{life_PL}}\right)$$

Proof of lemma (12):

$$\mathrm{IoR}_{\mathrm{life_PL}} = \mathrm{EL}^{\mathrm{EOP}}_{\mathrm{life_PL+NPL}} - \mathrm{EL}^{\mathrm{BOP}}_{\mathrm{life_PL+NPL}} + \mathrm{wo} =$$

$= \left(\mathrm{EL}_{\mathrm{PL+NPL}}^{\mathrm{EOP}} - \mathrm{EL}_{\mathrm{PL+NPL}}^{\mathrm{BOP}} + \mathrm{wo}\right) + \left(\mathrm{EL}_{\Delta\mathrm{PL}}^{\mathrm{EOP}} - \mathrm{EL}_{\Delta\mathrm{PL}}^{\mathrm{BOP}}\right) =$

$= \left(\mathrm{EL}_{\mathrm{PL}}^{\mathrm{EOP}} + \mathrm{PL\ Backtest} + \mathrm{NPL\ Backtest}\right) + \left(\mathrm{EL}_{\Delta\mathrm{PL}}^{\mathrm{EOP}} - \mathrm{EL}_{\Delta\mathrm{PL}}^{\mathrm{BOP}}\right) =$

$= \left(\mathrm{EL}_{\mathrm{PL}}^{\mathrm{BOP}} + \mathrm{PL\ Backtest} + \mathrm{NPL\ Backtest}\right) + \left(\mathrm{EL}_{\mathrm{life_PL}}^{\mathrm{EOP}} - \mathrm{EL}_{\mathrm{life_PL}}^{\mathrm{BOP}}\right)$ q.e.d.

$\mathrm{EL}_{\Delta\mathrm{PL}}^{\mathrm{BOP}}$ denotes the difference between (full or partially applied) lifetime EL and the regular 12-month EL. $\left(\mathrm{EL}_{\Delta\mathrm{PL}}^{\mathrm{EOP}} - \mathrm{EL}_{\Delta\mathrm{PL}}^{\mathrm{BOP}}\right)$ quantifies the excess conservativity during each accounting period introduced by life-time EL measures on performing portfolio compared to IRBA regulation. $\left(\mathrm{EL}_{\mathrm{life_PL}}^{\mathrm{EOP}} - \mathrm{EL}_{\mathrm{life_PL}}^{\mathrm{BOP}}\right)$ quantifies the excess conservativity compared to the neutral benchmark described in alternative 1. Both measures tend to be positive during growth periods and will become negative during liquidation.

Under both alternatives, the residual terms $\mathrm{PL\ Backtest} + \mathrm{NPL\ Backtest}$ remain unchanged, supporting an objective assessment of the quality of underlying EL parameters.

**4. Discussion of the PL Backtest**

(13) $\mathrm{EL}_{\mathrm{new\ NPL}}^{\mathrm{EOP}} + \mathrm{wo}_{\mathrm{new\ NPL}} = \mathrm{EL}_{\mathrm{PL}}^{\mathrm{BOP}} + \mathrm{PL\ Backtest}$

The PL Backtest is in its structure very similar to the traditional backtest for PD models. Instead of the traditional way of comparing unit-weighted default rates in homogenous segments, the PL Backtest compares PDs weighted by exposure at risk ($\mathrm{EAR}^{\mathrm{BOP}} = \mathrm{EAD}_{\mathrm{PL}}^{\mathrm{BOP}} * \mathrm{LGD}_{\mathrm{PL}}^{\mathrm{BOP}}$) at begin of period with the exposure at risk at end of period on non-performing customers ($\mathrm{EAR}^{\mathrm{EOP}} = \mathrm{EAD}_{\mathrm{NPL}}^{\mathrm{EOP}} * \mathrm{LGD}_{\mathrm{NPL}}^{\mathrm{EOP}}$).

Dividing (13) by ($\mathrm{EAD}_{\mathrm{PL}}^{\mathrm{BOP}} * \mathrm{LGD}_{\mathrm{PL}}^{\mathrm{BOP}}$) leads to:

(14) $\mathrm{rDF}_{\mathrm{BOP}}^{\mathrm{EOP}} = \mathrm{eDF}_{\mathrm{PL}}^{\mathrm{BOP}} + \mathrm{PD\ impact}$

where rDF is the (EAR- weighted) realized default frequency, and eDF the EAR-weighted portfolio average of PDs (expected default frequency). “PD impact” denotes the impact on eDF as measured by the PL Backtest. This notation outlines the similarity to the traditional PD backtest where unit-based RDFs (realized default frequencies) are compared to EDFs (expected default frequencies) without application of exposure-based weights.

A more business-oriented way is to express everything in risk densities (RD), dividing (13) by $\mathrm{EAD}_{\mathrm{PL}}^{\mathrm{BOP}}$:

(15) $\mathrm{rRD}_{\mathrm{BOP}}^{\mathrm{EOP}} = \mathrm{RD}_{\mathrm{PL}}^{\mathrm{BOP}} + \mathrm{RD\ impact}$

where rRD represents the EAD-weighted average net inflow into NPL (ie after application of the LGD), and “RD impact” denotes the impact of the PL Backtest on RD.

We conclude this section with the following observations:

a) A well established and stable default identification procedure is critical for the PL Backtest. In case of IRBA requirements, one needs to consider “likely to default” events consistently and implement 90 day past due not just in systems, but supporting collections processes have to be efficient to ensure a consistent and stable flow into default.

b) No assumption on the quality of the underlying PD model besides stability of the method is required: structural deviations (eg incorrect calibrations) will show up in the PL Backtest. The validity of the PL Backtest holds for expert based PD models as well as for highly sophisticated statistically derived ones. Only if the PD model per BOP was different to the one EOP, the change to $\mathrm{EL}_{\mathrm{PL}}^{\mathrm{BOP}}$ driven by the introduction of the new PD model has to be assessed separately. These changes are expected to be most relevant in case of significant re-calibrations or re-estimations of batch (behavior) PD models.

c) The LGD model definition changes from $\mathrm{LGD}_{\mathrm{PL}}$ in $\mathrm{EL}_{\mathrm{PL}}^{\mathrm{BOP}}$ (LGD method in performing portfolio) to $\mathrm{LGD}_{\mathrm{NPL}}$ in $\mathrm{EL}_{\mathrm{newNPL}}^{\mathrm{EOP}}$ (the LGD Best Estimate). Any discontinuities of the LGD model at default may distort the PL Backtest. For better understanding, the impact on $\mathrm{EL}_{\mathrm{newNPL}}^{\mathrm{EOP}}$ should be assessed separately by eg applying the $\mathrm{LGD}_{\mathrm{PL}}^{\mathrm{BOP}}$ in $\mathrm{EL}_{\mathrm{newNPL}}^{\mathrm{EOP}}$ for comparison purpose.

d) Regulators generally require PDs to be sufficiently conservative and to reflect long term average default rates (“through-the-cycle”, unconditional PDs). This requirement clearly causes a quantifiable bias in the PL Backtest driven by the conservativity requirement and the actually observed state of the economy in terms of “point-in-time” versus “through-the-cycle” risk levels. Assuming multiplicative relationship with factor (1+c) between “through-the-cycle”, conservative PDs and “point-in-time”, realized default rates in a certain year, the following can be derived:

(16) $\mathrm{rDF}_{\mathrm{BOP}}^{\mathrm{EOP}} = \mathrm{eDF}_{\mathrm{PL}}^{\mathrm{BOP}} * (1 + \mathrm{c}) = \mathrm{eDF}_{\mathrm{PL}}^{\mathrm{BOP}} + \mathrm{PD\ impact}$

(17) $\mathrm{PD\ impact} = -\,\mathrm{eDF}_{\mathrm{PL}}^{\mathrm{BOP}} * \mathrm{c}$, or

(18) $\mathrm{RD\ impact} = -\,\mathrm{eRD}_{\mathrm{PL}}^{\mathrm{BOP}} * \mathrm{c}$, or

(19) $\mathrm{PL\ Backtest} = -\,\mathrm{EL}_{\mathrm{PL}}^{\mathrm{BOP}} * \mathrm{c}$

Any imposed bias in the estimation of conditional, “point-in-time” PDs is linearly transferred to the PL Backtest and can be observed, verified and tested at this point. On the other hand, the PL backtest may be used to estimate c, ie the difference between conditional/point-in-time/unbiased and unconditional/through-the-cycle/conservative PDs.

Summarizing, a detailed discussion of the resulting PL Backtest needs to be focused on the identification of non-replicable outliers, replicable structural and/or cyclical impacts:

Non-replicable deviations may arise from big defaulting clients, caused by circumstances that may not repeat any more. Either they were caused by discontinued practices or other reasons, where non-replicability can be safely assumed.

Replicable deviations arise from model imperfections (biased models): conservativities as required to fulfill minimum regulatory standards were already discussed above. Further deviations may be caused by dependencies of PD with EAD (eg larger exposures defaulting more frequently than predicted by the PD) or simply by overly aggressive or conservative calibration of the underlying PD model.

Cyclical effects significantly impact the PL Backtest as well: during downturn periods with above-average losses the discussion of the PL Backtest necessarily needs to include an assessment whether excess defaults will repeat as the economical situation remains tough or whether normalization towards “through-the-cycle” EL will happen. In boom-times the opposite analysis has to be performed.

Typical segmentations to be investigated in the PL Backtest are portfolio splits by PD model (as there are different qualities of PD estimates to be assessed), by EAD or EAD times LGD (larger exposures may follow different underwriting routines or collections procedures than small exposures, causing obviously biased results) and PD itself (typically through an appropriate rating scale). Outliers (possible candidates for non-replicable defaults) may get identified by selecting the largest defaults sorted by $\mathrm{EL}_{\mathrm{newNPL}}^{\mathrm{EOP}}$. The PL Backtest is expected to be robust with respect to average and small defaults. In real life, few big defaults may explain a significant proportion of the PL Backtest deviation, depending on the level of concentration in a portfolio.

**5. Discussion of the NPL Backtest**

The NPL Backtest, as it appears in the decomposition of IoR, can be written as follows:

(20) $$\mathrm{EL}_{\mathrm{old\ NPL}}^{\mathrm{EOP}} + \mathrm{wo}_{\mathrm{old\ NPL}} = \mathrm{EL}_{\mathrm{NPL}}^{\mathrm{BOP}} + \text{ NPL Backtest}$$

It compares changes to the $\mathrm{EL}_{\mathrm{NPL}}^{\mathrm{BOP}}$ caused by actual recoveries, collateral releases or changes to collateral evaluations with the aging/slope structure of the $\mathrm{LGD}_{\mathrm{NPL}}$ model (EOP LGD is applied on the left hand side versus BOP LGD on the right hand side). In this sense it is the single-figure measure that compares the $\mathrm{LGD}_{\mathrm{NPL}}$ curve with cash flows and collateral changes during the observation period and quantifies any deviations.

The most significant bias in the NPL Backtest is introduced through the need to discount future cashflows in the LGD estimation by discount rate r. Implicitly this translates to a minimum earned income of r percent per year on the net balance (ER = EAD – EL, also denoted as “expected discounted recoveries” or non-performing net receivables). The impact of this discounting requirement on the NPL Backtest is $-\mathrm{r}\left(\mathrm{EAD}_{\mathrm{NPL}}^{\mathrm{BOP}}-\mathrm{EL}_{\mathrm{NPL}}^{\mathrm{BOP}}\right)= -\mathrm{r}\,\mathrm{ER}_{\mathrm{NPL}}^{\mathrm{BOP}}$, as shown in the appendix under question 2.

Conservativity, downturn or indirect costs may be modeled in many different ways depending on the specific structure of the $\mathrm{LGD}_{\mathrm{NPL}}$ model in a business. For simplicity, it is assumed that any further conservativity requirements are modeled through a (most likely higher) discount rate $r^c$ applied in the LGD estimation. The total impact on the NPL Backtest becomes:

(21) $$-\mathrm{r}^{\mathrm{c}}\left(\mathrm{EAD}_{\mathrm{NPL}}^{\mathrm{BOP}}-\mathrm{EL}_{\mathrm{NPL}}^{\mathrm{BOP}}\right)= -\mathrm{r}^{\mathrm{c}}\,\mathrm{ER}_{\mathrm{NPL}}^{\mathrm{BOP}}$$

For further reading on (conservative) Downturn LGD versus LGD Best Estimate considerations please follow up with question 3 in the appendix.

We conclude this section with the following observations:

a) The need for conservative LGD estimates under a downturn scenario (formally another conservativity requirement) and the need to discount future recoveries lead to a structurally (negative) expected deviation in the NPL Backtest. Although these deviations cannot be avoided for regulatory reasons, its quantification and discussion is required for internal steering.

b) The slope of $\mathrm{LGD}_{\mathrm{NPL}}$ (evolution by time, speed of appreciation towards 100%) has strong impact on the result of the NPL Backtest. If the slope is too steep initially, recoveries (or cures) may not be sufficient to reflect this slope and a shortage may be reported in the NPL Backtest. If the slope is too flat, the write-off may lead to a shortage later if $\mathrm{LGD}_{\mathrm{NPL}}$ does not approach 100% quickly enough. As long as write-offs occur regularly, the LGD model is challenged effectively by comparing its prediction with actual account development during the observation period.

Comment: there might be local regulations or business practices which allow or require very late write-off. In that case, overly flat LGD slopes (especially if the LGD remains significantly below 100% for long time) may remain undetected. To cover this possible gap, the speed of decline of the stock of non-performing net receivables should be measured as well:

(22) $$\mathrm{ER}_{\mathrm{old\,NPL}}^{\mathrm{EOP}}=\mathrm{ER}_{\mathrm{NPL}}^{\mathrm{BOP}}+\ \mathrm{RecoFlow}$$

Or

(23) $\text{RecoFlow} = \text{ER}_{\text{old NPL}}^{\text{EOP}} - \text{ER}_{\text{NPL}}^{\text{BOP}}$

RecoFlow is a measure for the ongoing effectiveness of collection efforts on non-performing loans. It quantifies cash payments or proceeds from collateral repossessions, which reduce the overall stock of non-performing net receivables. Risk mitigation activities like additionally provided guarantees or collateral increase non-performing net receivables and have an adverse effect on RecoFlow. These risk-mitigating activities, delaying risk recognition by increasing the stock of non-performing net receivables now, must finally translate into cash proceedings and positive RecoFlow performance.

Remark: the sum of (20) and (22) actually describes the EAD movement for the NPL portfolio:

(24) = (20) + (22) $\text{EAD}_{\text{old NPL}}^{\text{EOP}} + \text{wo}_{\text{old NPL}} = \text{EAD}_{\text{NPL}}^{\text{BOP}} + \text{ NPL Backtest} + \text{RecoFlow}$

c) The LGD model definition may change during the observation period (model refinements or re-estimations). In that case, the new LGD model should be simulated per BOP to separate the impact from the model change in the NPL Backtest to derive a meaningful result for the new LGD model.

Similar to the PL Backtest, the discussion of the NPL Backtest needs to be focused on the identification of non-replicable, replicable structural and/or cyclical impacts:

Non-replicable deviations may arise from untypically large recoveries or revaluations of collateral (eg collateral lost or not accessible, vanished in value), caused by circumstances that may not repeat any more. Either they were caused by discontinued practices or other reasons, where non-replicability can be safely assumed.

Replicable deviations arise from model imperfections: required conservativities or downturn effects were already discussed above. Discounting of future recoveries leads to another formal deviation, although a clear requirement for fair value estimation. Further deviations may be caused by dependencies of LGD with EAD (ie larger exposures have different recovery streams than small ones), PD or simply by a too steep or too flat slope of the underlying $\text{LGD}_{\text{NPL}}$ model.

Finally cyclical effects significantly impact the NPL Backtest: during downturn periods with below-average recoveries and reduced collateral evaluations, the discussion of the NPL Backtest necessarily needs to include an assessment whether these impacts will continue as the economical situation remains tough or whether normalization towards a more regular “through-the-cycle” scenario can be expected.

The key segmentations to be investigated in the NPL Backtest are by LGD model segments (different qualities of LGD estimates), exposure (different sizes may be treated differently in collection procedures), collateral type and time in default (to get better information on the alignment of the slope of $\mathrm{LGD}_{\mathrm{NPL}}$ with recoveries by year in default). Outliers (possible candidates for non-replicable recoveries) are identified by ranking the NPL Backtest results – large positive (recovery shortage, collateral revaluation, steep LGD slope) and large negative deviations (cured or repaid accounts) indicate possible candidates. Few large cases may explain a significant proportion of the NPL Backtest deviation in real applications, depending on the level of concentration in a portfolio.

## 6. The relationship between the PL/NPL Backtest and traditional backtesting in IRBA validations

To align the proposed concept to IRBA calibration techniques, the PL Backtest can be split into the following three components, separating PD, EAD and LGD impacts:

(25) $$\text{PL Backtest} = \mathrm{EL}^{\mathrm{EOP}}_{\text{new NPL}} + \mathrm{wo}_{\text{new NPL}} - \mathrm{EL}^{\mathrm{BOP}}_{\mathrm{PL}} =$$

$$= \sum_{\text{new Defaults}} \left(\mathrm{PD}^{\mathrm{EOP}}\mathrm{EAD}^{\mathrm{EOP}}\ \mathrm{LGD}^{\mathrm{EOP}}_{\mathrm{NPL}} + \mathrm{wo}\right) - \mathrm{EL}^{\mathrm{BOP}}_{\mathrm{PL}} =$$

$$= \sum_{\text{new Defaults}} \left(\mathrm{PD}^{\mathrm{EOP}}\mathrm{EAD}^{\mathrm{EOP}}\ \mathrm{LGD}^{\mathrm{EOP}}_{\mathrm{NPL}} + \mathrm{wo} - \mathrm{PD}^{\mathrm{DEF}}\mathrm{EAD}^{\mathrm{DEF}}\ \mathrm{LGD}^{\mathrm{DEF}}_{\mathrm{PL}}\right) +$$

$$+ \sum_{\text{new Defaults}} \left(\mathrm{PD}^{\mathrm{DEF}}\mathrm{EAD}^{\mathrm{DEF}}\ \mathrm{LGD}^{\mathrm{DEF}}_{\mathrm{PL}} - \mathrm{PD}^{\mathrm{DEF}}\mathrm{EAD}^{\mathrm{BOP}}\ \mathrm{LGD}^{\mathrm{BOP}}_{\mathrm{PL}}\right) +$$

$$+ \sum_{\text{new Defaults}} \mathrm{PD}^{\mathrm{DEF}}\mathrm{EAD}^{\mathrm{BOP}}\ \mathrm{LGD}^{\mathrm{BOP}}_{\mathrm{PL}} - \mathrm{EL}^{\mathrm{BOP}}_{\mathrm{PL}} =$$

$$= \Delta_{\mathrm{LGD}} + \Delta_{\mathrm{EAD}} + \Delta_{\mathrm{PD}}$$

DEF stands for the respective value at time of default. For a fair comparison with IRBA validation concepts, $\mathrm{PD}^{\mathrm{DEF}}$ and $\mathrm{PD}^{\mathrm{EOP}}$ can be assumed to be 100%, since the impact of PD-reducing guarantees on non-performing clients is typically considered separately. Similarly $\mathrm{EL}^{\mathrm{BOP}}_{\mathrm{PL}}$ should be calculated before application of PD-reducing guarantees for the same reason.

Under this assumption,

(26) $$\Delta_{\mathrm{PD}} = \textstyle\sum_{\text{new Defaults}} \mathrm{EAD}^{\mathrm{BOP}}\ \mathrm{LGD}^{\mathrm{BOP}}_{\mathrm{PL}} - \mathrm{EL}^{\mathrm{BOP}}_{\mathrm{PL}}$$

backtests estimated to realized defaults, weighted by $\mathrm{EAD}^{\mathrm{BOP}}\ \mathrm{LGD}^{\mathrm{BOP}}_{\mathrm{PL}}$. In comparison, IRBA calibration tests focus on (unit-based) average PDs and realized default rates. Under the IRBA calibration assumption of homogenous segments (specifically with respect to the level of $\mathrm{EAD}^{\mathrm{BOP}}\ \mathrm{LGD}^{\mathrm{BOP}}_{\mathrm{PL}}$), both concepts are essentially equivalent.

(27) $$\Delta_{\mathrm{EAD}} = \textstyle\sum_{\text{new Defaults}} \left(\mathrm{EAD}^{\mathrm{DEF}}\ \mathrm{LGD}^{\mathrm{DEF}}_{\mathrm{PL}} - \mathrm{EAD}^{\mathrm{BOP}}\ \mathrm{LGD}^{\mathrm{BOP}}_{\mathrm{PL}}\right)$$

is comparing realized EADs at default with the EAD estimate at BOP, weighted by $\mathrm{LGD}_{\mathrm{PL}}^{\mathrm{BOP}}$. Instead of unit-based backtest-approaches under IRBA, these weights ensure proper differentiation by clients with higher or lower economic loss expectations. As before, under the assumption of homogenous segments with respect to $\mathrm{LGD}_{\mathrm{PL}}^{\mathrm{BOP}}$, both approaches are equivalent.

Comment: The usage of $\mathrm{LGD}_{\mathrm{PL}}^{\mathrm{DEF}}$ as weight for $\mathrm{EAD}^{\mathrm{DEF}}$ is required to properly adjust for collateral repossessions just before default. Without this adjustment, repossessions may offset exposure increases before default, leading to a conservative EAD backtest which is not materially justified.

These LGD discontinuities are quantified by

(28) $$\Delta_{\mathrm{LGD}} = \sum\nolimits_{\text{new Defaults}}\left(\mathrm{EAD}^{\mathrm{EOP}}\ \mathrm{LGD}_{\mathrm{NPL}}^{\mathrm{EOP}} + \mathrm{wo} - \mathrm{EAD}^{\mathrm{DEF}}\ \mathrm{LGD}_{\mathrm{PL}}^{\mathrm{DEF}}\right)$$

In the typical case that $\mathrm{LGD}_{\mathrm{NPL}}$ at day 1 after default is identical (or very close to) to the $\mathrm{LGD}_{\mathrm{PL}}$ and continuously increasing later, the observed difference should be small. As before, collateral repossessions (which might also indicate a discontinuity in LGD, but a materially justified one) are properly accounted for due to the respective application of $\mathrm{EAD}^{\mathrm{EOP}}$ and $\mathrm{EAD}^{\mathrm{DEF}}$.

(29) $$\text{NPL Backtest} = \mathrm{EL}_{\text{old NPL}}^{\mathrm{EOP}} + \mathrm{wo}_{\text{old NPL}} - \mathrm{EL}_{\mathrm{NPL}}^{\mathrm{BOP}}$$

finally compares the (lifetime) economic loss estimation at begin of period with the sum of one year incremental economic gain/loss (deviation of collection performance to LGD prediction) and the economic loss estimate at end of period (again a lifetime estimate, but 1 year replaced by actual performance). It is like taking one specific slice of a vintage analysis and folding them such that the overall LGD performance in a specific year can be observed. There is no comparable counterpart in IRBA validations, where LGD backtests typically are based on long-term vintage analyses to compare LGD estimates with realized discounted cash flows.

For improved transparency, the NPL Backtest should be split by years in default, highlighting changing LGD performance in different time periods after default. Early high cure rates may offset poor late workout performance as long as only the overall NPL Backtest is considered.

Under the assumption that different assumptions underlying internal LGD estimates cause a significant part of the unexplained RWA deviations currently observed for otherwise comparable portfolios, the NPL Backtest (together with RecoFlow) may become a powerful tool to objectively assess LGD model quality compared to actual business performance and to benchmark deviations across different entities.

Summarizing, the PL and NPL Backtests can be aligned to traditional IRBA validation concepts, and allow for a structured and clear view on the overall and combined performance of EL parameters. Due to its reconciliation with reported EL measures and capital consumption through "Impact of Risk", the resulting deviations observed in the PL and NPL Backtests are highly relevant for parameters in use for internal steering, such as pricing or the estimation of economic capital.

## 7. Discussion of the $\mathrm{EL}_{\mathrm{PL}}^{\mathrm{EOP}}$

After adjusting Impact of Risk (IoR) for deviations observed in the PL Backtest and the NPL Backtest, the remaining term is properly described by the $\mathrm{EL}_{\mathrm{PL}}^{\mathrm{EOP}}$. There is not much to be told – the $\mathrm{EL}_{\mathrm{PL}}^{\mathrm{EOP}}$ can best be compared to the insurance premium for credit risk. Whenever a credit portfolio is generated based on new volume or updated PDs are assigned to a portfolio, the resulting $\mathrm{EL}_{\mathrm{PL}}^{\mathrm{EOP}}$ is immediately incurred through capital requirement. The shortfall adjustment in IRBA regulation effectively ensures this relationship. Now the decomposition of IoR is complete:

$\mathbf{EL}_{\mathbf{PL}}^{\mathbf{EOP}}$, the standard measure for credit risk.

**PL Backtest**, measuring point-in-time deviations in actual default performance and is a good starting point to discuss outliers, economical cycles and its impact on default rates or other structural deviations. It may highlight calibration biases in the PD relevant for internal steering applications.

**NPL Backtest**, measuring actual point-in-time recovery performance, compares it to the expectation as modeled by the LGD and is a good starting point to discuss outliers, economical cycle impacts or other structural deviations. It may highlight a calibration biase in the LGD relevant for internal steering applications.

**RecoFlow**, measuring the ongoing decrease of expected recoveries, ie the net receivables stock of non-performing loans, as antipode to the NPL backtest with respect to EAD.

PL Backtest, NPL Backtest and RecoFlow serve as effective credit risk indicators to assess the calibration of EL compared to flow into default, collection efficiency on non-performing portfolio and final write-off amounts. In case of ongoing, replicable, structural biases the EL can be adjusted for these deviations in internal steering applications.

## 8. How the proposed EL Backtest might prove useful in implementation of IFRS 9:

The following comment is mostly based on result of the EFRAG (2013) field test, feedback by the Basel committee (2013) and Deloitte (2013) on the IFRS 9 exposure draft 2013/3, and on recent publications addressing RWA inconsistencies under Basel 2 regulation by eg Samuels (2013), Cannata *et al* (2012) and PECDC (2013). It reflects a personal and subjective opinion by the author, which does not necessarily correspond to the opinion of UniCredit Bank Austria AG. It may require significant analytical or empirical follow up depending on the way IFRS 9 finally develops.

The following three critical areas of IFRS 9 can be identified, where the application of adequately applied EL backtesting methods may help to reduce inconsistencies and provide objective benchmarks. The identified areas are lifetime EL estimation and verification, the threshold between bucket 1 and bucket 2 and the point-in-time property of estimates.

a) Lifetime EL estimation and verification:

IFRS 9 assumes the availability of high quality, predictive point-in-time lifetime EL estimates across the whole credit portfolio. This is in many directions a significantly stronger methodological requirement than imposed by Basel 2: IRBA may be limited to specific segments where methodological developments are feasible, whereas IFRS 9 has to be applied to the whole credit portfolio. Lifetime EL requires an extended (lifetime) predictive horizon, which under IRBA is only required for LGD. Specifically, the estimation of remaining maturity as part of lifetime EL estimation may prove challenging. Estimates need to be calibrated point-in-time, whereas IRBA through-the-cycle calibrations should be based on statistically stable long term averages. And predictions by economists, which are supposed to be considered in point-in-time estimates, increase subjectivity and methodological complexity.

There are already a few years experience from IRBA implementations since rollout of Basel 2 available. Although banks under IRBA have to capture significant amount of historic data (covering well-defined minimum time series requirements), have to align parameters with business processes (through the use-test requirements) and are heavily regulated and annually validated (by different independent bodies such as internal validation, internal audit and external regulators, to name a few), significant unexplained differences between resulting RWAs across otherwise comparable portfolios can still be observed.

IFRS 9 on the other hand does not contain detailed guidance with respect to maturity or lifetime EL, nor does it establish validation concepts to ensure consistent estimation.

The proposed PL/NPL Backtests based on expected loss could provide a valuable tool to be included in a validation framework under IFRS 9, with some adjustments to reflect lifetime EL and the 3-bucket approach: One way of integrating lifetime EL measures was outlined in section 3. Another one is to replace the PL-NPL split by the threshold between bucket 1 and bucket 2, and to replace the LGD in NPL portfolio by lifetime EL in bucket 2 and bucket 3. Nevertheless, significant methodological guidance is expected from IASB to ensure consistency across all entities.

b) The threshold between bucket 1 and bucket 2:

This threshold takes a similar role as the default identification under Basel 2 regulation and will become the most significant discontinuity ("cliff") in the provisioning process under IFRS 9. Considering that default identification under Basel 2 is highly regulated (detailed definitions of thresholds, days overdue, types of unlikely-to-pay events, discussions on loss of present value), the guidance under IFRS 9 is limited ("significant increase in credit risk") and opens a wide range for different interpretations by entity and segment.

The suggested PL Backtest, applied on monthly data for improved granularity, is an effective tool to validate the consistency of the default timeseries under Basel 2. For IFRS 9, a similar validation for the flow between bucket 1 and bucket 2 can be defined to verify consistency between entities and segments and to monitor the flow in different periods of the economic cycle.

c) The point-in-time requirement for estimates and procyclicality:

Current research for downturn LGD is frequently based on market implied LGDs, a concept mostly relevant for investment portfolios. For commercial lending, longer collections periods tend to mask short-term downturn impacts in overall estimates. With bucket 2 in IFRS 9 however, cyclical effects currently mostly observed in PD will become increasingly relevant for provisioning. As immediate consequence, the procyclical tendency of provision generation/release will increase compared to IAS 39. This might add structural risk to the sector: during good economic conditions entities are required to release reserves. Only when evidence for a downturn appears (either through changed economic forecasts or through shocks like Lehman Brothers), provisions have to be generated. Although the overall level of provisions will be higher than under IAS 9 even under good economic conditions, this cliff effect is more severe and has the potential to cause more problems for entities than ever before, ie actually increasing default risk in the sector.

This risk could be mitigated by developing earlier warning indicators for economic downturns, although history proves that this lead time generally is relatively short.

For an analytical approach to PIT parameter estimation, an identification of the actual (historical) portfolio status (ie loss performance) with respect to TTC conditions is required. Due to the high volatility inherent in the loss distribution for typically concentrated commercial portfolios, this is a statistically challenging task. On basis of the proposed framework (specifically the PL Backtest) a maximum-likelihood estimator for the external factor in the loss distribution model on basis of monthly data points might be developed to derive an analytically founded PIT/TTC concept. It needs to be noted, however, that this task requires significant further research and empirical analyses.

The question remains, whether the industry should focus capacities to predict actual downturns – containing significant uncertainty – or to build robust risk management processes more in line with the antifragility concept of Taleb (2012). Under the current draft of IFRS 9 it would be easy to release reserves during years of favorable economic conditions (driven by strong business pressure, supported by accounting standards) and to distribute available capital to shareholders. Whether a possible deterioration of economical conditions can be detected early enough to generate the funds necessary for the requested accelerated provisioning (eg by retaining profits) remains to be proven. In worst case, entities might end up short of provisions in case of sudden economic changes. These increased methodological and structural risks introduced by IFRS 9 must be managed as carefully as possible.

## 9. Summary

A new framework was proposed to align parameter based EL estimates with actual Impact of Risk (IoR). For this purpose, a more general view on credit risk was developed, adding capital charges through shortfall to the traditional P&L impacts of credit risk. Deviations between $\mathrm{EL}_{\mathrm{PL}}$ and IoR can be linked to PD and/or LGD by means of the newly introduced concepts of the PL Backtest and the NPL Backtest. The framework is applicable under a wide range of parameter qualities (expert driven up to sophisticated and statistically derived), as long as default identification procedures and credit risk parameters remain stable.

This calculus is just a starting point: it outlines the methodological basics necessary to understand the practical properties of EL. It has obvious consequences for pricing (eg by correcting overly aggressive or conservative EL-based risk margin estimates) and other risk-adjusted profitability measures for financial institutions (eg RAROC or EVA). It has the additional potential to become a valuable tool to verify and backtest lifetime EL measures planned to be used for IRFS 9 provisioning.

## 10. Appendix

Question 1: Is overall risk quantity increased or decreased by observing Impact of Risk (IoR) instead of provision- (P&L-) based measures?

Assume a portfolio generated in year 1 and being 100% liquidated (paid back or written-off) at year n. $\mathrm{EL}^0$ , $\mathrm{EL}^1$ to $\mathrm{EL}^n$ are measured at the end of each year with $\mathrm{EL}^0 = \mathrm{EL}^n = 0$. $\mathrm{wo}^1$ to $\mathrm{wo}^n$ representing the write-off amounts in years 1 to n, respectively. The following can be derived:

(30) $\sum_1^n \mathrm{IoR}^i = \sum_1^n\left(\mathrm{EL}^i - \mathrm{EL}^{i-1} + \mathrm{wo}^i\right) = \sum_1^n \mathrm{wo}^i$

The same can be proven for provision based risk measures (Cost of Risk, CoR):

(31) $\sum_1^n \mathrm{CoR}^i = \sum_1^n\left(\mathrm{LLP}^i - \mathrm{LLP}^{i-1} + \mathrm{wo}^i\right) = \sum_1^n \mathrm{wo}^i$

It can be seen that total risk measures are identical (up to discounting effects) as long as a full cycle of a portfolio from generation until final liquidation is considered, independent of the specific measurement. Differences during specific accounting periods are driven by timing of default, the amount of conservativity included in the EL estimate, provisioning and write-off processes. Even complete or partial application of lifetime, point-in-time EL instead of EL in equation (30) would not change this property.

For a simplified but illustrative example take a synthetic portfolio of 10.000 contracts, each of size 1 unit, generated in June year 1, with term 1 year, 100% bullet capital payment after 12 monthly interest installments. 200 contracts (PD=2%) will actually default in year 2, out of which 100 (50%) can be collected in year 3. Write-off of the remaining 100 (50%=LGD) happens in year 4. Net risk margin as charged is 1%, earned partly in year 1 (50 units) and year 2 (50 units).

Case 1, correctly estimated parameters (PD, EAD and LGD), unbiased EL:

EL by end of year 1, when all accounts are booked, is 100 units (2% PD at 50% LGD). Consequently, IoR in 1st year is 100 units, twice the earned net risk margin. Impact on capital is incurred slightly front-loaded under IRBA (partly through IBNR, partly through remaining shortfall), on average 6 months before it gets earned by the customer.

In year 2 200 defaults are observed. EL of new defaults ($= \mathrm{EL}_{\mathrm{newNPL}}^{\mathrm{EOP}}$) in year 2 is 100 units (=200*50% LGD), IoR becomes 0 and the PL Backtest closes at 0 as well.

In year 3, 50% recoveries are received while the LGD increases to 100%: $\mathrm{EL}_{\mathrm{NPL}}^{\mathrm{EOP}}$ remains at 100 units (200-100 recoveries) at 100% LGD), IoR remains 0 and the NPL Backtest closes at 0 as well (remark: no discounting of future cash flows is considered in this example).

In year 4, the remaining 100 units have to be written-off), IoR remains 0 and the NPL backtest closes at 0 again. Summarizing: In year 1, the full impact of risk of 100 units is incurred, without any further impact later on. Both backtests have perfect result in each period, with 0 deviation.

<u>Case 2</u>, PD estimate is too aggressive at 1%, LGD remains at 50%:

EL by end of year 1 is 50 units, IoR becomes 50, 50% of what was observed in case 1.

When 200 defaults occur in year 2, the EL of new defaults ($\mathrm{EL}_{\mathrm{newNPL}}^{\mathrm{EOP}}$) in year 2 becomes 100 units, (200*50%), leading to a shortage of 50 in the PL Backtest result, caused by the aggressive PD estimate. IoR catches up: another 50 can be observed, the total IoR is now identical to case 1.

In year 3 and 4 the situation is unchanged. Summarizing: Total impact of risk again is 100 units, but incurred with a delay of up to 12 months. The delay is effectively observable in the PL Backtest.

<u>Case 3</u>, PD at correct 2%, LGD too aggressive at 25%, after receiving recoveries increases to 50%:

EL by end of year 1 is again 50 units, IoR remains 50.

When 200 defaults occur in year 2, the PL Backtest compares EL end of year 1 (50 units) with EL of new defaults ($\mathrm{EL}_{\mathrm{newNPL}}^{\mathrm{EOP}}$) in year 2, also 50 units (200*25%), leading to a perfect PL Backtest result since the same, overly aggressive LGD is applied on both sides of the backtest. IoR is 0 in year 2.

In year 3, 50% recoveries are received and the LGD on the remaining balance increases to 50%: $\mathrm{EL}_{\mathrm{NPL}}^{\mathrm{BOP}}$ = 50 compared to $\mathrm{EL}_{\mathrm{NPL}}^{\mathrm{EOP}}$ of (200-100 recoveries)*50% LGD = 50. Again IoR remains 0 and the NPL Backtest still closes in at 0 as well.

In year 4, when finally the remaining 100 units have to be written-off, the NPL Backtest compares $\mathrm{EL}_{\mathrm{NPL}}^{\mathrm{BOP}}$ of 50 with 100 write-off, showing a shortage of 50. Observed IoR is 50, and the total IoR adds up to 100 units as in case 1 and 2.

Summarizing: Total EL starts at 50 in year 1 and stays at this level until year 4, when finally – through write-offs – the full extent of risk impact on capital will be incurred. IoR is 50 in year 1 and 4, zero in between. The delay is effectively observable in the NPL Backtest.

Imperfect parameters impact the timing of risk impact on capital – aggressive PD estimates may delay risk recognition by one year whereas aggressive LGD estimates may delay risk recognition until final write-off. The total cumulative risk impact on capital however remains unchanged, independent of the quality of parameter estimates.

Question 2: How to estimate the impact of discounting on NPL Backtest?

Lemma (32) $\text{NPL Backtest} = -r\left(\text{EAD}_{\text{NPL}}^{\text{BOP}} - \text{EL}_{\text{NPL}}^{\text{BOP}}\right)$

Proof of lemma (32): Under the assumption of correctly predicted recovery cash flows per year,

$$\text{EL}_{\text{NPL}}^{\text{BOP}} = \text{EAD}_{\text{NPL}}^{\text{BOP}} - \text{Rec}^1 \frac{1}{(1+r)} - \text{Rec}^2 \frac{1}{(1+r)^2} - \text{Rec}^3 \frac{1}{(1+r)^3} - \cdots$$

and

$$\text{EL}_{\text{oldNPL}}^{\text{EOP}} = \text{EAD}_{\text{oldNPL}}^{\text{EOP}} - \text{Rec}^2 \frac{1}{(1+r)} - \text{Rec}^3 \frac{1}{(1+r)^2} - \cdots$$

hold, where r is the rate used for discounting of cash flows in the LGD estimation. $\text{Rec}^i$ are recoveries (absolute values) received at the end of year i for the portfolio under consideration, net of direct expenses charged on the account. Now the NPL Backtest becomes:

$$\text{NPL Backtest} = \text{EL}_{\text{old NPL}}^{\text{EOP}} + \text{wo}_{\text{old NPL}} - \text{EL}_{\text{NPL}}^{\text{BOP}} =$$
$$= \text{EAD}_{\text{oldNPL}}^{\text{EOP}} - \text{Rec}^2 \frac{1}{(1+r)} - \text{Rec}^3 \frac{1}{(1+r)^2} - \cdots$$
$$- \text{EAD}_{\text{NPL}}^{\text{BOP}} + \text{Rec}^1 \frac{1}{(1+r)} + \text{Rec}^2 \frac{1}{(1+r)^2} + \text{Rec}^3 \frac{1}{(1+r)^3} + \cdots$$
$$+ \text{wo}_{\text{old NPL}} =$$
$$= \text{EAD}_{\text{oldNPL}}^{\text{EOP}} - \text{EAD}_{\text{NPL}}^{\text{BOP}} + \text{Rec}^1 \frac{1}{(1+r)} + \text{wo}_{\text{old NPL}} -$$
$$-\text{Rec}^2 \frac{1+r-1}{(1+r)^2} - \text{Rec}^3 \frac{1+r-1}{(1+r)^3} - \cdots =$$
$$= \text{EAD}_{\text{oldNPL}}^{\text{EOP}} - \text{EAD}_{\text{NPL}}^{\text{BOP}} + \text{Rec}_1 + \text{wo}_{\text{old NPL}} -$$
$$-r * \left(\text{Rec}^1 \frac{1}{(1+r)} + \text{Rec}^2 \frac{1}{(1+r)^2} + \text{Rec}^3 \frac{1}{(1+r)^3} + \cdots\right) =$$

$$= -r * \left(\text{Rec}^1 \frac{1}{(1+r)} + \text{Rec}^2 \frac{1}{(1+r)^2} + \text{Rec}^3 \frac{1}{(1+r)^3} + \cdots\right) =$$
$$= -r\left(\text{EAD}_{\text{NPL}}^{\text{BOP}} - \text{EL}_{\text{NPL}}^{\text{BOP}}\right)$$
q.e.d.

$\text{EAD}_{\text{oldNPL}}^{\text{EOP}} = \left(\text{EAD}_{\text{NPL}}^{\text{BOP}} - \text{Rec}_1 - \text{wo}_{\text{old NPL}}\right)$ under the assumption that there is no further change in EAD parameters during the observation period.

Question 3: Is there anything to be considerd regarding specific properties of LGD/EL Best Estimate?

Formally, in some regulations the $\mathrm{LGD_{NPL}}$ (LGD Best Estimate or EL Best Estimate) does not necessarily need to be conservative nor needs to reflect downturn scenarios (reflecting "point-in-time"). However, any differences to the conservative downturn LGD lead to additional capital requirement by increasing RWA through $12{,}5*(\mathrm{EL}-\mathrm{LGD_{NPL}})$ in IRBA. It is materially irrelevant for impact on capital whether the (conservative downturn) LGD leads to incremental shortfall or whether the difference between the downturn LGD and the $\mathrm{LGD_{NPL}}$ increases RWA. Consequently, to keep the framework simple, it can be assumed that $\mathrm{LGD_{NPL}}$ is consistent with downturn LGD with regards to conservativity requirements.

Some biographical notes:

After a period of academic research at the Institute for Econometrics at the University of Technology in Vienna with Prof. Dr. Manfred Deistler, I spend almost 20 years in banking area mostly working on data based analysis of credit risk. Starting with scoring development and provisioning methods along with US-GAAP regulations, I later became deeply exposed to methodical accounting-related questions such as income accrual and other P&L analysis. With the advent of Basel II, I returned into parameter based credit risk modeling and validation, contributing significantly to the success of the IRBA implementation of UniCredit Bank Austria AG. After taking over responsibility on all credit risk parameters PD, EAD and LGD and on basis of my experience in accounting, P&L and balance sheet dynamics, the presented calculus of expected loss was developed.